# Evaluating the robustness of top coatings comprising plasma-deposited fluorocarbons in electrowetting systems

Dimitrios P. Papageorgiou, Elias P. Koumoulos, Costas A. Charitidis,

Andreas G. Boudouvis and Athanasios G. Papathanasiou[*]

*School of Chemical Engineering, National Technical University of Athens, GR-15780, Athens, Greece*

Short title: **Evaluating the robustness of top coatings in electrowetting**

[*]Author to whom correspondence should be addressed. Tel.:+30-2107723290; e-mail: pathan@chemeng.ntua.gr

**Abstract**

Thin dielectric stacks comprising a main insulating layer and a hydrophobic top coating are commonly used in low voltage electrowetting systems. However, in most cases, thin dielectrics fail to endure persistent electrowetting testing at high voltages, namely beyond the saturation onset, as electrolysis indicates dielectric failure. Careful sample inspection via optical microscopy revealed possible local delamination of the top coating under high electric fields. Thus, improvement of the adhesion strength of the hydrophobic top coating to the main dielectric is attempted through a plasma-deposited fluorocarbon interlayer. Interestingly enough the proposed dielectric stack exhibited a) resistance to dielectric breakdown, b) higher contact angle modulation range, and c) electrowetting cycle reversibility. Appearance of electrolysis in the saturation regime is inhibited, suggesting the use of this hydrophobic dielectric stack for the design of more efficient electrowetting systems. The possible causes of the improved performance are investigated by nanoscratch characterization.

1. **Introduction**

Electrowetting (EW) deals with the enhancement of the wetting properties of solids by the modification of the electric charge density at a liquid/solid interface. Suitable application of external electric field induces variation of the contact angle of conductive liquids on insulating substrates such as polymers, glass and oxides. EW can provide more than 100° of contact angle modulation reversibly, especially in an oil ambient, with fast response to actuation in the order of milliseconds [1]. As a result, EW has been utilized for a number of technological applications such as lab-on-chip devices [2, 3], liquid lenses [4, 5], electronic displays [6, 7] and "smart" microbatteries [8], to name a few. For all these applications it is desirable to use low voltages to induce contact angle changes, through either reduction of the dielectric thickness or the use of ionic surfactants [9]. However, indication of dielectric failure (most commonly electrolysis) is frequent, especially in cases of thin dielectrics. Consequently, improving the robustness of the dielectric is of great importance since it is related to the robustness of devices.

Electrowetting on dielectric (EWOD, usually called EW) can be realized when a conductive sessile drop sits on a hydrophobic dielectric on top of a conductive electrode [10]. The dependence of the contact angle, $\theta_V$, on the applied voltage, $V$, is given by the Lippmann equation [11],

$$\cos\theta_V = \cos\theta_Y + \frac{1}{2\gamma}CV^2, \qquad (1)$$

$$C = \varepsilon_0\varepsilon_r/d,$$

where $\theta_Y$ is the Young's contact angle and $\gamma$ is the liquid surface tension. $C$ is the capacitance per unit area, $d$ is the thickness of the dielectric with dielectric constant $\varepsilon_r$, and $\varepsilon_0$ is the permittivity of vacuum. Lippmann equation demonstrates reliable predictions of $\theta_V$ at low voltages, however at high voltages experiments show that, beyond a critical voltage, $V_s$, the contact angle (CA) reaches a lower limit in contradiction to Eq. (1) which predicts complete wetting, i.e. $\theta_V=0°$ at sufficiently high applied voltage. This phenomenon is widely known as CA saturation that limits the EW response to the applied voltage. Recent studies attribute the CA saturation to leakage current mechanisms, i.e. dielectric breakdown [12], dielectric charge trapping [13, 14] and air ionization [15] caused by the increased electric field strength in the vicinity of the three-phase contact line (TPL). Material breakdown at the onset of saturation coupled with the charge leakage propagating through the dielectric is of great research importance, as the understanding of the related mechanisms could lead to more efficient EW devices.

A usual choice for hydrophobic dielectric is stand-alone amorphous fluoropolymers (FPs). Stand-alone FPs inherently feature high porosity of the polymeric compound. The pores are considered to be conductive paths, which could lead more easily to dielectric breakdown in EW experiments especially when thin FP films are used. High porosity of the FP films is related to substantial charge injection; pore density affects charge injection which is related to limited EW response to the applied voltage. For this purpose, superior dielectrics in terms of insulation properties are used [e.g. $SiO_2$, $Si_3N_4$, tetraethoxysilane (TEOS)] which have lower porous density than FPs, thus dielectric breakdown is suppressed.

Since EW can only reduce CA, the chosen dielectric should be hydrophobic so as to achieve the highest CA modulation range possible. As a result, it is evident that a hydrophobic dielectric stack, namely a main insulating layer and a hydrophobic top coating, is requisite to attain high CA modulation and improved resistance to dielectric breakdown. This approach gives rise to separate optimization of the dielectrics and the hydrophobic coatings for the benefit of the EW response to the applied voltage.

In the case of dielectric optimization, stacked main dielectrics, such as $SiO_2$ - $Si_3N_4$ - $SiO_2$ [oxide-nitride-oxide (ONO)] have been investigated extensively for use in integrated circuit memories. It was found that this type of structure exhibits higher breakdown field than the conventional single layer dielectrics (i.e. $SiO_2$, $Si_3N_4$). Tested ONO samples in our group [16] showed roughly 25% higher saturation voltage and almost 10° higher CA modulation, than the equivalent $SiO_2$ dielectric.

FPs as hydrophobic top coatings are an intriguing part of the hydrophobic dielectric investigated. In addition to the commercially available FPs (Teflon®, Cytop® and Fluoropel®), plasma-deposited fluorocarbons (FCs) were used as hydrophobic top coatings, which feature several distinct advantages [17]. Plasma technology is still under investigation for EW applications. Also self-assembled monolayers (SAMs) are used as hydrophobic coatings, however, the CA electrowetting irreversibility limits SAMs applicability in devices [18].

The main objective of this study was to improve the performance of the hydrophobic dielectric in terms of resistance to dielectric breakdown in EW tests and provide an estimation of the top coating adhesion strength to the oxide substrates tested.

Experiments in our group showed that a possible cause of EW degraded response might be the inadequate adhesion of the hydrophobic top coating to the main insulating substrate, coupled with the high porosity of the top coating material. Usually during EW sample testing, for applied voltages higher than $V_s$ (saturation voltage), bubbles rise inside the liquid drop signaling electrolysis and ultimately sample failure. Detailed post inspection of the sample surface showed that the top coating starts to fail, in the form of random surface micro-cracks. Moreover, in some cases local delamination of the top coating from the substrate was noticeable. To study the potential of improving the adhesion strength of the top coating to the substrate and to reduce the porosity aspects of the hydrophobic dielectric, a plasma-deposited FC layer was implemented [19, 20]. EW experiments were conducted on this proposed hydrophobic dielectric stack as well as qualitative evaluation of the top coating adhesion to the main dielectric was attempted, through nanoscratch tests.

Nanoscratch testing is a versatile tool for analysis of the mechanical properties of thin films and bulk materials. In scratch tests, a diamond stylus is drawn over the film surface under progressively increasing normal load (NL) until the film is detached from the substrate. Single scratches with a ramped NL are useful for critical load ($L_c$), film adhesion and mar studies. The critical load ($L_c$), corresponding to film delamination, can provide a measure of the scratch resistance or adhesion strength of the film, but it is difficult to extract adhesion strength quantitatively since the critical load depends not only on adhesion strength but also on several intrinsic and extrinsic factors. While the intrinsic factors are related to the test conditions such as loading rate, scratching speed, and indenter shape, the extrinsic parameters are connected to film–substrate system such as material properties, friction coefficient and physical dimensions [21]. That is why, all adhesion measurement techniques (including scratch test) are generally considered to measure what is called "practical adhesion" [22]. Though various models have been developed to obtain cohesive strength of the film and adhesion strength between the film and the substrate via conventional indentation procedure, there is no standard methodology for quantitative assessment.

Reported herein is a sandwich-like hydrophobic top coating comprising a plasma-deposited fluorocarbon and a spin coated fluoropolymer on top of TEOS. EW tests on this proposed composite top coating showed resistance to dielectric breakdown, reversible EW behavior and improved adhesion strength, compared to other FP coatings tested. Comparison of the adhesion strength (adhesion to the substrate) between the composite coating and FP coatings commonly used in EW experiments was qualitatively assessed through nanoscratch tests. Our objective was to estimate whether the interlayer mechanical properties were a key factor in EW device design.

**2. Materials and Methods**

Various hydrophobic dielectric stacks were fabricated on phosphorus-doped Si wafers which were also used as ground electrodes (resistivity, 1-10 $\Omega$/cm). The hydrophobic dielectric stacks consist of a main dielectric and a hydrophobic top coating. $SiO_2$ or TEOS were used as the main dielectrics. Commercial amorphous fluoropolymers (AFs) such as Asahi Cytop® 809M, Teflon® AF 1600 and plasma-deposited fluorocarbons (FCs) were used as hydrophobic top coatings.

The adhesion strength of Teflon® to various substrates is most commonly improved with the use of silanes. As a result, fluorosilanes are used as primers for the Teflon® AF [23] and, in particular, perfluorooctyltriethoxysilane solution is spin coated onto the oxide layer and the coated wafers are heated at 95 °C for 15 min. Teflon® AF is then spun on top of the fluorosilane layer.

Asahi Cytop® 809M, as a commercial AF alternative, is diluted in perfluorohexene and spun on top of $SiO_2$ (35 nm thick Cytop®). A special process sequence, in an oven, is needed for the Cytop to adhere well to the oxide surface.

In this work, on top of TEOS, an alternative hydrophobic top coating was used. A thin plasma FC film (30-100 nm) was deposited as an adhesion promoter layer for the commercial Teflon® AF [19]. Teflon® AF (30-60 nm) was diluted at Fluorinert® Fluid FC-77 solvent; and then spin coated on the plasma FC film. After spinning, the sample was baked in air at 95°C for 5 min.

Verification of the thicknesses of the oxide and the top coating layers was performed with a spectroscopic ellipsometer, model M2000 J.A. Woolam Co. (accuracy in the measured thickness ± 0.5 nm).

AC Electrowetting (2.3 kHz sine wave) measurements are conducted in oil ambient. The samples are immersed in a completely transparent poly(methyl methacrylate) (PMMA) oblong tank filled with 99+% pure dodecane. The sessile droplet consists of 0.1% sodium dodecyl sulfate (SDS) in 0.1N NaCl (conductivity ~ 11.22 mS/cm).

Measurements of the dependence of the CA on the applied voltage were performed in an in-house built EW experimental setup, previously described in Papathanasiou et al. [16]. Real time image processing software, that was developed in-house, was used to analyze the drop shape. The method is described in [16] and the accuracy is of the order of ±1.5°.

The surface of the hydrophobic top coatings was inspected in detail with an optical microscope (Zeiss AX10 Imager.A1m). Immediately after the EW experiments the sessile drop was removed from the sample for optical characterization of the drop's footprint.

Nanoscratch testing was performed with Hysitron TriboLab® Nanomechanical Test Instrument, which allows the application of loads from 1 to 10.000 μN and records the displacement dependence on applied load with high load (1 nN) and a high displacement (0.04 nm) resolution. The TriboLab® employed in this study is equipped with a Scanning Probe Microscope (SPM), in which a sharp probe tip moves in a raster scan pattern across the sample surface using a three-axis piezo positioner. All nanoscratch measurements were performed with the standard three-sided pyramidal Berkovich probe, with an average radius of curvature of about 100 nm, in a clean area environment with 45% humidity and 23°C ambient temperature [24].

The scratch tests performed in this work included three main segments. Firstly, a pre-scratch scan under a very small load (1 μN) was carried out. Then, the indenter scraped the sample under a certain force and a scratch was generated. The normal applied loads (NL) used in this work were 50-300 µN. The length of the scratches was 10 µm. Finally, a post-scratch test under the same NL as the pre-scratch test was conducted to obtain the image of the surface after scratch. An estimation of the residual scratch ditch and the extent of immediate recovery can be obtained by comparing the pre-scratch with the post-scratch image profiles.

## 3. Results and Discussion

### 3.1 Electrowetting on Composite Hydrophobic Coating

In this work we focused on the investigation of adequate coupling in terms of interlayer adhesion strength and chemical affinity of the hydrophobic dielectric used in EW experiments, namely the main insulating layer (TEOS) and the hydrophobic top coating. There are a number of known issues (i.e. dielectric charging, electrolysis) related to hydrophobic dielectrics in EW systems, which can either suppress the CA modulation range or cause sample failure. The investigation of the CA modulation at voltages $V > V_s$ is limited by the fact that dielectric breakdown is most likely to occur. Bubbles rise inside the liquid drop indicating electrolysis and ultimately sample failure. Detailed post inspection of the top coating (Teflon®) showed that under high electric fields it could possibly locally delaminate from the substrate. This observation led to a thorough investigation of alternatives to enhance the adhesion between the hydrophobic top coating (Teflon®) and the main dielectric (TEOS).

The adhesion of Teflon® to substrates (e.g. silicon, glass) depends primarily on physical interaction since it has no reactive chemical groups for chemical bonding [19]. Fluorosilanes, that were originally used in our group to promote adhesion between Teflon® coating and TEOS, proved to be inadequate for investigating the electrowetting CA modulation at voltages higher than $V_s$. Electrolysis was still present during the experiments in the saturation regime (at $V_s$ and beyond). Our study showed that the adhesion of Teflon® AF to TEOS could be improved by the use of a thin plasma-deposited FC layer. Plasma-deposited FC films are known to adhere well to oxide surfaces due to an oxyfluoride interface layer on which a Teflon-like (1 < F/C < 2) layer grows (F/C stands for "fluorocarbon ratio") [25]. It is the chemical affinity of the plasma-deposited FC to Teflon® that improves the overall bondability of Teflon® AF to the oxide substrate. The result is a sandwich-like hydrophobic coating, hereafter called "Composite Coating", which consists of a thin plasma-deposited FC layer and a thin spin coated Teflon® film.

In Figure 1, EW experiments on the tested samples are presented. The EW tests were performed in dodecane ambient as follows: The applied voltage was increased from 0 Volts in increments of 2.5 Volts up to the critical voltage, namely $V_s$, where CA saturation sets on. Then the voltage was turned off and the sessile droplet rested back in its initial shape. This will be from now on referred to as an EW cycle. Moreover, robustness verification in terms of dielectric breakdown prevention was performed. For this purpose, composite coated samples were compared to Teflon® coated ones with respect to the CA dependence on applied voltages up to $2.5V_s$.

Usually, EW experimental data are presented up to the saturation limit, and compared with the predictions of Young-Lippmann equation. In rare cases and for relatively thick hydrophobic dielectrics, experimental data for applied voltages $V > V_s$ are presented [26]. In this work three samples were tested at applied voltages apparently beyond the saturation. The samples consist of TEOS as the main dielectric (with thicknesses of 180 nm and 821 nm) and on top of it the following hydrophobic top coatings were fabricated: Two composite coatings (with thicknesses of 58 nm and 174 nm) and one Teflon® coating (52 nm thick).

As expected, the experimental data are in close agreement with the predictions of Young-Lippmann equation (dashed lines in Fig.1a, b) up to the onset of saturation. In Fig.1a, samples of equal TEOS thicknesses and different types of top coatings (composite and Teflon® coatings) are compared, in terms of CA dependence on the applied voltage. For an applied voltage of 15V, the contact angle modulation is 110° (from ~160° to ~50°) for both samples. However, the EW tests show that the Teflon® coated sample failed at about $1.4V_s$ (see arrow in Fig.1a), as bubbles start to emerge from the sample surface. In contrast, the composite coating sample performs a CA of 37° at 44.2V and a consequent maximum CA modulation of ~125°.

The robustness of this composite coating sample was tested at high voltages, up to $2.5V_s$. We did not notice any sample failure indication (e.g. electrolysis) during this test, however, CA modulation gradually decreases from a maximum of 125° to 110°. This gradual loss of performance is evident up to the fourth EW cycle and thereafter CA modulation range remains constant for at least up to thirty cycles.

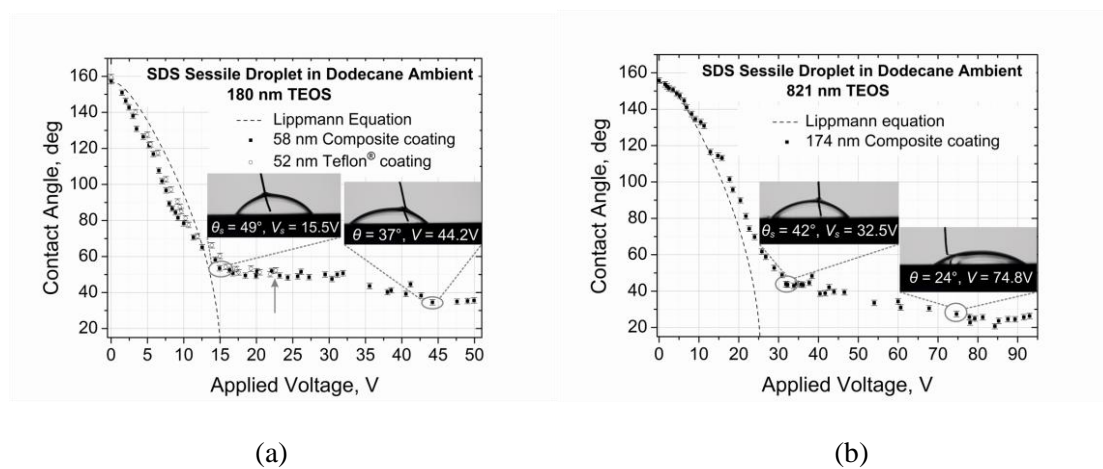

(a)                                                                  (b)

**Figure 1.** Contact angle dependence on the applied voltage, (a) comparison between the composite and Teflon® coatings, (b) for the composite coating.

In Fig. 1b, the performance of a composite coated sample is shown in terms of maximum CA modulation. For an applied voltage of 74.8V the maximum contact angle modulation is ~130°. Wetting enhancement of 130° can be achieved at $2.5V_s$, without material failure. Material failure is also not evident at least up to thirty EW cycles. Similarly, CA modulation decreased to 115° during this EW test. Gradual decrease in maximum CA modulation range is observed in all composite coated samples tested at high voltages. Possibly charge trapping in the hydrophobic dielectric suppresses the wetting enhancement during the EW cycle sequence [14]. It should be mentioned that CA hysteresis (difference between the advancing and the receding CA) was ~5°.

**3.2 Optical Microscopy Characterization of the Hydrophobic top Coatings**

The surface of the hydrophobic coating was inspected by optical microscopy, immediately after the EW tests. The objective here was to examine the footprint, namely the effect of the EW test on the surface of the coating, of the electrowetted sessile drop for different hydrophobic top coatings. Optical observation was focused in the vicinity of TPL, because the electric field strength is expected to reach very high values, thus greater surface damage is expected [27].

The samples were carefully inspected after the following EW experimental procedure: The applied voltage was increased stepwise up to $2.5V_s$ and was held constant for about 30 s at each EW experiment. Then the voltage is turned off and the above cycle is repeated at least up to thirty times. The sessile drop is then removed and the sample surface is inspected with optical microscopy. The specifications of each tested sample featuring 300nm TEOS as the main dielectric, are presented in Table 1.

**Table 1.** Specifications of the samples inspected.

| Sample | Coating | Coating Thickness (nm) | In Air Static CA Hysteresis (deg) | In Dodecane (SDS droplet) EWOD Hysteresis (deg) | Saturation Voltage, $V_s$ (Volts) |
|---|---|---|---|---|---|
| S1 | Plasma FC | 60 | 42±2.5 | - | 18±1 |
| S2 | Teflon® | 60 | 11±2.5 | 5±2.5 | 17.8±1 |
| S3 | Composite | 60 | 12±2.5 | 5±2.5 | 17.3±1 |

Three samples that consist of different hydrophobic top coatings were fabricated (see Table 1). The first sample (sample S1) that features only plasma-deposited FC coating on top of TEOS shows resistance to dielectric breakdown. However, upon voltage removal, the sessile drop does not recede to its initial shape and stays at its advanced wetting state. Static CA hysteresis of the sessile drop is 42°, which indicates high EW irreversibility [28]. Since it was not possible to perform reversible EW cycles due to high hysteresis, application of voltage for a long time was decided to test the robustness of the sample at high voltages. Our experiments showed that even if a voltage of the order of 2.5$V_s$ was applied for 5 minutes, there was no sign of electrolysis. The microscopy inspection of the sample surface shows noticeable damage (Fig. 2a); however, material breakdown is not evident in the EW test. Clearly in the vicinity of the TPL there is a narrow band (~80 μm) that suggests that this portion of the surface is mostly affected. The stressed area looks like a ring with a narrow band at the edge, formed by the fully advanced wetting state of the drop. Although we observe these random formations, there is no macroscopic indication of material damage (i.e. electrolysis) that usually happens on Teflon® coating which will be discussed below.

The second sample (sample S2), with Teflon® AF as hydrophobic coating, appears highly affected (Fig. 2b), at the edge of the ring and on the inside of it. At applied voltage ~1.7$V_s$, bubbles started to rise inside the liquid drop indicating electrolysis. The narrow band at the edge of the ring is clearly visible in this sample (~75 μm). Inside the ring, random tree-like formations possibly indicate the presence of extended surface cracks. Similarly, Cytop® as hydrophobic top coating failed at applied voltage ~1.4$V_s$. Detailed inspection of the surface reveals that Cytop® coating exhibits a ripple-like topography which can be possibly attributed to material structure different from the Teflon® coating, however, there is no significant damage or crack formation to report. Comparison of the figures 2a and 2b shows that the corresponding materials, namely plasma FC and Teflon® AF, behave differently under high electric fields.

The third sample tested (sample S3) features the proposed composite hydrophobic top coating (Fig. 2c). This coating combines the advantages of the previous top coatings as it shows resistance to sample failure and reversibility in EW tests. The footprint of the drop is more uniform and the ring in the vicinity of the TPL is still evident in all tested samples, with a narrow band of ~50 μm. It should be mentioned that the sample remains fully functional at least up to thirty EW cycles. The inspection of the surface revealed the absence of dendritic patterns seen in sample S2, although the upper layer is the same i.e. Teflon® AF.

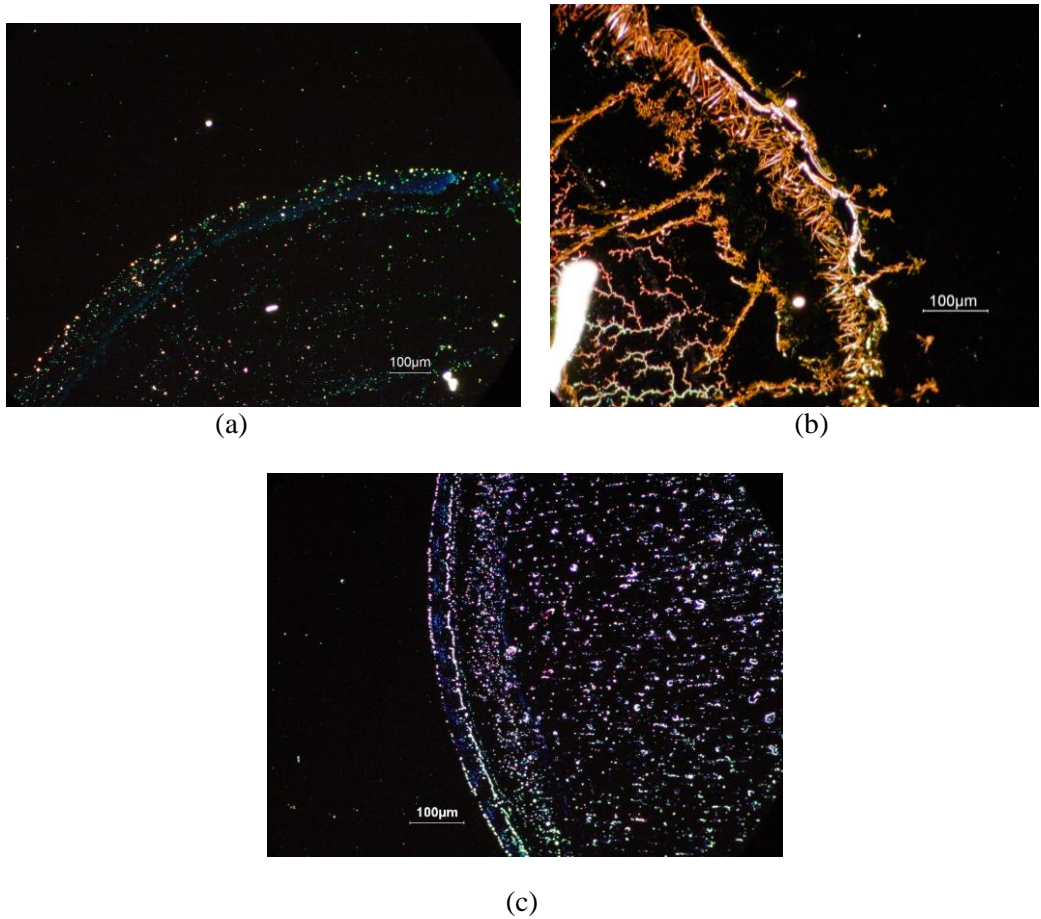

**Figure 2**. The effect of the EW tests on sample surfaces for a) Plasma FC coating, b) Teflon® coating and c) Composite coating.

It should be noted that when electrolysis happens, bubbles are localized in the vicinity of the TPL, confirming the high electric field strength in this region. The plasma-deposited FC interlayer might have a twofold advantage: on the one hand reduced void density between the hydrophobic coating and TEOS through better adhesion and on the other hand inhibition of local charge trapping in the overall hydrophobic top coating through reduced porosity. In the following section we focus on the interlayer mechanical properties of the hydrophobic dielectric to estimate the contribution of this factor to the overall EW system performance.

.

**3.3 Nanoscratch tests**

Nanoscratch tests can provide a measure of the scratch resistance of the hydrophobic dielectric. Initially, nanoindentation tests were conducted to determine the hardness and elastic modulus of hydrophobic dielectric layers. The corresponding values for each layer were used to determine the sequence parameters for the following scratch tests, i.e. applied normal load (NL), scratch length, tip velocity. Two samples were tested. Sample S3 consists of 300 nm TEOS and 60 nm composite top coating, namely, 30 nm spin coated Teflon® on top of 30 nm plasma-deposited FC; Sample S2 consists of 300 nm TEOS and 60 nm spin coated Teflon®.

The pre-scratch scan curve corresponds to the profile of the initial flat surface, scratch scan curve corresponds to the tip penetration profile during testing and finally, post-scratch scan curve corresponds to the final profile of the surface after scratching (1 µN). The post-scratch curve represents the residual depth of scratch trace after unloading, i.e. the plastic deformation of the probed film. The statistical error in the scratch depths is less than 10 nm, while the displacement resolution of the nanoindenter used in scratch testing is better than 0.1 nm, thus making possible to compare the scratch depths of different films.

The difference between the scratch and post-scratch curves corresponds to the elastic recovery of the films, making nanoscratch testing a reliable technique for defining the elastic and plastic regions of thin coatings. The surface profiles of the scratches of the two samples under 50 µN of NL and the subsequent initial (scratch scan) and residual (post-scratch scan) scratch depth profiles are presented in Fig. 3.

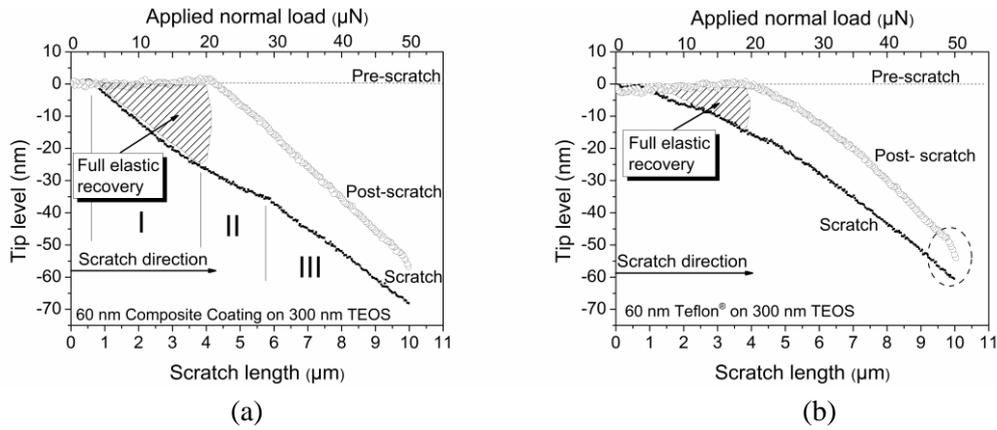

**Figure 3.** Scratch profiles for the a) composite coating sample (sample S3) and b) Teflon® coating (sample S2) on top of TEOS, with applied NL of 50 μN.

The three regimes in Fig. 3a (I, II, III) are defined by the different slopes of the scratch scan curve. In detail, for sample S3 (Fig. 3a) and for NL lower than 18 μN (regime I), the penetration depth of the indenter did not exceed 30 nm. This fact could be correlated to the mechanical response of the top layer (Teflon®) and the underlying plasma FC layer of the composite coating. For sample S2 (Fig. 3b) and for NL lower than 18 μN the penetration depth of the indenter was ~20nm which is solely attributed to the mechanical properties of the Teflon® coating. For NL higher than 18 μN and up to 30 μN (regime II), the lower slope of the curve indicates that the plasma FC layer contributes to the overall scratch resistance of the composite coating to the applied NL [29]. Moreover, from the structural material viewpoint, the plasma FC interlayer is expected to be more resistant to the NL than the Teflon® AF. The structure (-C-C-) that these polymers are consisted of is related to material hardness [30]. Chemical characterization of the plasma FC films through composition (XPS) analysis has shown that the plasma-deposited F/C ratio is 1.5 [31], whereas the F/C ratio of Teflon is 2. Hence, plasma-deposited FC is more crosslinked than the Teflon® AF (more (-C-C-) bonds per volume). As a result, it is not surprising that the plasma-deposited FC appears to be more resistant to the NL.

Comparing the surface profiles of Figures 3a,b, up to 20 μN of applied NL, the recovered scratch depth of the sample S3 is close to 30 nm, whereas the recovered scratch depth of the sample S2 is about 15 nm (hatched areas in Figs. 3a and b). The higher elastic recovery under the same NL suggests that the composite coating can sustain higher scratch induced stresses (higher scratch resistance, i.e. improved adhesion strength) than the Teflon[®] coating [29]. After NL ~30μN (regime III) both samples exhibited elastoplastic behaviour, with sample S2 exhibiting almost full plastic behavior (convergence of the initial and the residual scratch profiles) in the last few nanometers of displacement (indicated in Fig. 3b with a dashed circle).

In Figure 4 the surface profiles of the two scratched samples are shown. As seen in Figs. 4a and b, there is a buildup of polymer material mostly on one side of the scratch [32]. These buildups are found in all scratches created, which shows that the films were plastically deformed and that the buildup was most likely an accumulation of compressed materials.

When a moving scratch tip ploughs through the coating, the material will be either pushed forward or piled-up sideways ahead of the tip; material's pile-up on the sides of the indenter suggests plastic deformation of a film over an undeformable substrate [33, 34]. This phenomenon is usually observed for relatively ductile polymers, where plastic deformation is evident on applied strains [35]. In Fig. 4c cross-sectional scratch profiles (via SPM imaging at maximum load) of samples S3 and S2 at 50,150 and 300 μN of applied NL are presented.

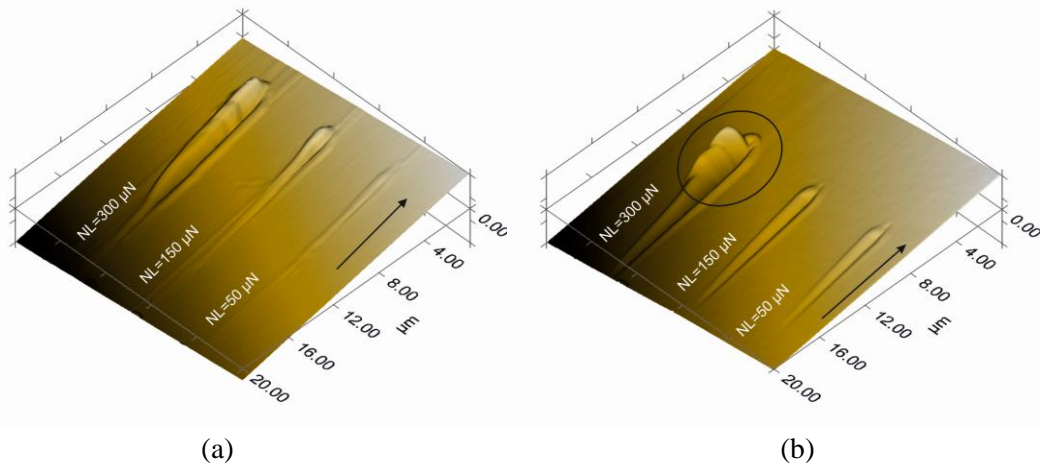

(a)                          (b)

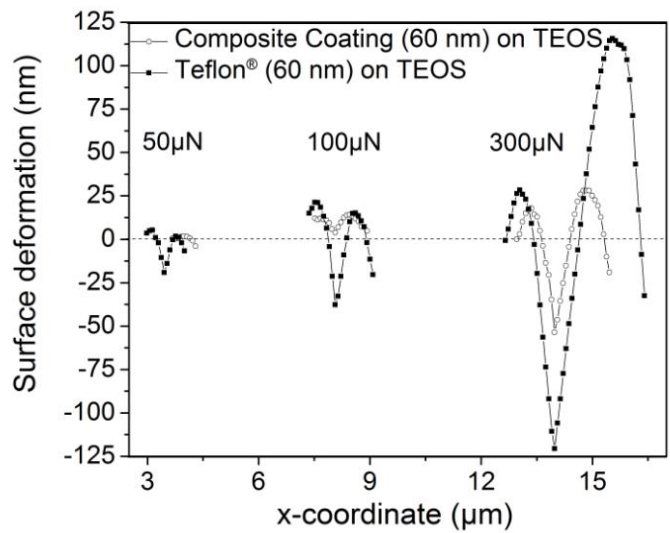

(c)

**Figure 4.** Scratched surface profiles: SPM images at 50 μN, 100 μN and 300 μN of applied NL for a) composite coating sample (sample S3), b) Teflon® coated sample (sample S2). c) Cross sectional shape at maximum applied NL of 50 μN, 100 μN and 300 μN, i.e. at the end of scratch. Pointing arrow defines the scratch direction of the tip. (Zero level in the vertical axis corresponds to the level of the unscratched flat surface)

At high nanoscratch loads (300 μN) for the Teflon® coated sample (sample S2), the polymeric material accumulated on the sides of the tip which is apparent from Figs. 4b and c. The result, shown in Fig.4b, is an indication of coating failure, which resulted in a blister sample damage (indicated by the circle in Fig. 4b). The displacement of the removed material on the sides of the scratch, indicates that scratching caused mainly plastic deformation. This type of deformation of coating is possibly attributed to buckling and delamination effects, observed in the coating when it is subjected to scratch test. Significant pile-up of the material due to diamond tip scratching reveals wear and debris. In the case of the composite coating sample (sample S3) the ploughing by the tip does not create similar accumulation pattern (Figs. 4a and c), and consequently coating failure is suppressed. It should be noted though that sample S2 exhibited earlier accumulation of material. Larger plastic deformation in the case of sample S2 compared to sample S3 is evident by the residual depths, since in both cases the total penetration depth is almost identical.

In every Teflon® coated sample tested by nanoscratching, we observed an abrupt change in NL at a certain scratch depth (ranging from 68 to 76 nm). This abrupt change is attributed to a discontinuity of certain mechanical properties (e.g. elastic modulus) between Teflon® and TEOS which reflects to the adhesion between the two and consequently an observed abrupt change in applied NL. Since this abrupt change was never observed in the case of the composite coating sample, it is suggested that plasma-deposited FC smooths out the aforementioned discontinuity between TEOS and the hydrophobic top coating (Teflon®), resulting in better EW performance.

In Figure 5, applied NL dependences on scratch depth and length are presented. The tip scratches the surface under progressively increasing NL and along a predefined path. As denoted by the arrow, an abrupt change in NL is observed only when Teflon® coating is used, which is indicative of strength weakening due to material heterogeneity [21, 29, 36]; this occurs at a critical scratch length of ~5.5 μm and NL ~ 150 μN.

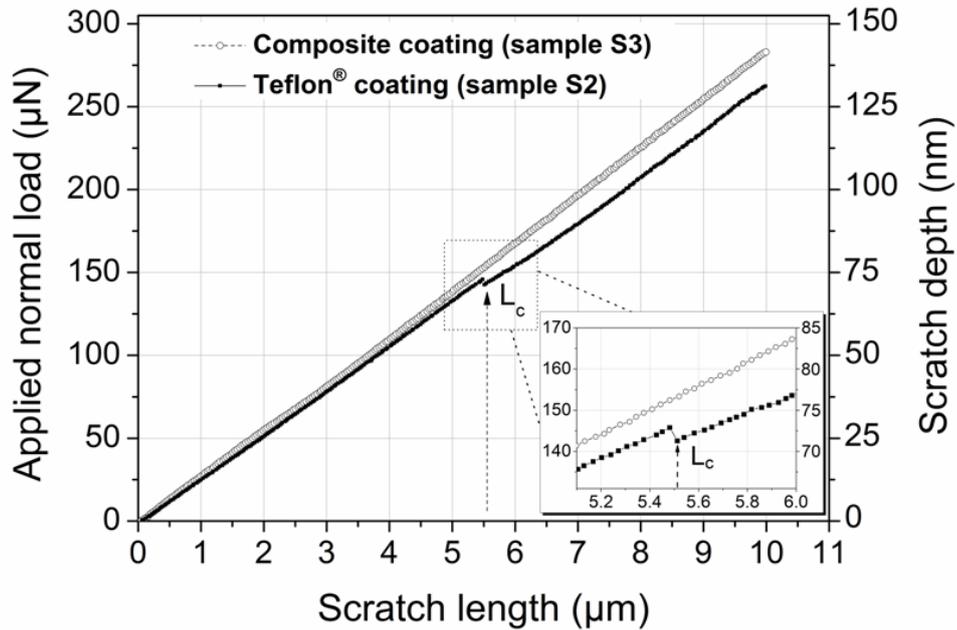

**Figure 5.** Applied normal load dependences on the scratch length and scratch depth for samples S3 and S2. The arrow indicates the onset of critical load. The inset is a magnification of the figure area depicted by the dashed line.

The scratch depth variation ( see Fig. 5), indicates that the NL abrupt change sets in when the tip penetration is close to the Teflon®/TEOS interface (sample S2). The corresponding critical load is usually denoted as $L_c$. The existence of an $L_c$ is an indication of failure in terms of coating cracking, delamination or brittle fracture caused by scratch testing [37]. High elasticity in combination with low hardness of the Teflon® top coating mostly favor delamination and not coating cracking or brittle fracture. Moreover, the Teflon® coating of the tested sample is approximately 60 nm thick which is close to the scratch depth value where the critical load appears. This strain mismatch evident by the abrupt change in the applied NL induces film delamination, and is not observed in the case of the composite coating. We suspect that the interlayer of plasma FC suitably bonds the oxide substrate and the spin coated Teflon® layer, therefore the corresponding nanoscratch curve in Fig. 5 is smoother for the composite coating.

**Conclusions**

In this work the effect of plasma-deposited fluorocarbons, as structural layers of the top coating, on EW performance was investigated. A sandwich-like hydrophobic top coating was fabricated, here called composite coating, comprising a thin plasma-deposited FC layer and a thin spin coated Teflon® layer. This sample showed resistance to dielectric breakdown, improved CA modulation and reversibility for at least up to thirty EW cycles, at applied voltages apparently beyond the saturation. Optical microscopy inspection revealed absence of dendritic patterns usually observed in Teflon® coatings. Nanoscrach testing was conducted to further investigate the interlayer mechanical properties of the proposed hydrophobic dielectric stack. Nanoscratch measurements showed improved adhesion strength of the composite coating to the oxide substrate compared to the equivalent Teflon® coating sample, confirming the observed improved robustness in EW tests.

**Acknowledgements**

The research leading to these results received funding from the European Research Council under the European Community's Seventh Framework Programme (FP7/2007-2013) / ERC Grant agreement n° [240710]. The authors wish to thank Dr E. Gogolides and Dr A. Tserepi at the Institute of Microelectronics, NCSR "Demokritos", for their expert advice in plasma-deposited fluorocarbon films and clean-room processing. The authors also wish to thank Dr Panagiota Petrou, and Dr Sotirios Kakabakos from the Institute of Radioisotopes and Radiodiagnostic Products of NCSR "Demokritos" for providing access to their optical microscope. The authors would like to thank Prof. T. Krupenkin and Dr. J. Ashley Taylor of the University of Wisconsin-Madison for kindly providing samples with Cytop® hydrophobic top coating.